\documentclass[preprint,review,12pt]{elsarticle}

\usepackage{float}
\usepackage{amsmath}
\usepackage{subcaption}
\usepackage{graphicx}
\usepackage{tabularx}
\usepackage{siunitx}

\DeclareSIUnit\bar{bar}
\DeclareSIUnit\angstrom{\text{A°}}

\usepackage[dvipsnames]{xcolor}

\usepackage{url}

\usepackage{enumitem}
\usepackage{amssymb}

\usepackage[figuresright]{rotating}
\usepackage{lineno}
\journal{Nuclear Physics A}

\begin{document}

\begin{frontmatter}

\title{Ionization quenching factors and $W$-values of low-energy H$_2^+$ and He$^+$ ions in Ar gas at low pressure measured with a bulk resistive MICROMEGAS}

\author[2,1]{A. Foresi \corref{cor1}}
\author[1]{G. Antonelli}
\author[1]{C. Avanzini}
\author[1]{G. Balestri}
\author[2,1]{G. Bigongiari} 
\author[1]{E. Bossini}
\author[4]{M.G. Callaini}
\author[1]{R. Carosi}
\author[3]{E. De Angelis}
\author[1]{F. Frasconi}
\author[2,1]{P. Maestro \corref{cor1}} 
\author[1]{F. Morsani}
\author[3]{A. Mura}
\author[1]{L. Orsini}
\author[1]{G. Petragnani}
\author[1]{F. Pilo \corref{cor1}}
\author[3]{R. Rispoli}
\author[1]{G. Terreni}
\cortext[cor1]{Corresponding authors: foresi@pi.infn.it, maestro@unisi.it, pilo@pi.infn.it}
\affiliation[1]{organization={INFN Sezione di Pisa},
addressline={Polo Fibonacci, Largo B. Pontecorvo 3},
postcode={56127},
city={Pisa},
country={Italy}}
\affiliation[2]{organization={Department of Physical Sciences, Earth and Environment, University of Siena},
addressline={via Roma 56},
city={Siena},
postcode={53100},
country={Italy}}
\affiliation[3]{
organization={INAF-IAPS Istituto Nazionale di Astrofisica - Istituto di Astrofisica e Planetologia Spaziali},
city={Rome},
country={Italy}}
\affiliation[4]{organization={Department of Physics, University of Pisa},
addressline={Largo Bruno Pontecorvo 3},
city={Pisa},
postcode={56127},
country={Italy}}

\begin{abstract}
The ionization quenching factor, the fraction of an ion's initial kinetic energy lost through ionization in a medium,
was measured for H$_2^+$ and He$^+$ ions within the  2.5-5 keV energy range 
in an Ar/CO$_2$ gas mixture at pressures between 75 and 150 mbar. 
The mixture was contained in the active volume of a MICROMEGAS type Micro Pattern Gaseous Detector (MPGD), 
which was connected to an ion source via a high vacuum system.
The results showed a significant decrease in ionization as the ion energy decreased, although no dependence on gas pressure was observed within the low pressure range studied. Additionally, significant deviations from the predictions of SRIM (Stopping and Range of Ions in Matter) simulations were found.
These measurements also allowed the first determination of $W$ values (the average energy required to create an electron-ion pair in the gas) for H$_2^+$ and He$^+$ ions in Ar at this particular energy range, which had not been previously explored for these projectile-target combinations.
These findings are essential for accurate ion energy reconstruction in low-pressure MPGDs, 
with particular relevance for space-based energetic neutral atom detection,  as proposed by the Italian Space Agency's SWEATERS project. 
\end{abstract}
\end{frontmatter}
\section{Introduction}
Accurate measurement of the ionisation caused by low-energy ions in a medium is essential for calibrating the energy response of detectors used in several fields, including microdosimetry \cite{microdos}, direct dark matter detection \cite{DDMD_MIMAC_Santos_2011_TrackReconstruction}, plasma diagnostics \cite{ITER_NPASystems_2010} and space weather monitoring.
In the latter field, detectors sensitive to low-energy 
Energetic Neutral Atoms (ENAs) have been developed \citep{SpaceWeather_TWINS_McComas_2009, Fuselier, Orsini} and used in solar physics and planetary exploration missions to study the dynamics of planetary magnetospheres and the evolution of magnetic perturbations caused by solar flares and coronal mass ejections. 
In planetary environments, ENAs are produced by charge-exchange interactions between solar wind and magnetospheric ions with neutral atoms of the planetary exosphere. 
These newly formed ENAs retain the energy of the original ions. 
On the Earth, ENA energies are generally limited to below a few hundred keV, constrained by the ion population energy distribution and the energy dependence of the differential cross section \cite{DetectionOfENAsPWurz}. 
Because ENAs are unaffected by Earth's geomagnetic field, remote sensing from low Earth orbit (LEO) satellites offers a valuable method for imaging various regions of the magnetosphere, such as the ring current and the bow shock \citep{SpaceWeather_Gruntman_ENAReview_1997, Gold, Krimigis, Galli}. These regions exhibit distinct ENA populations and energy spectra, with the bow shock predominantly emitting hydrogen ENAs at around 2 keV, 
with a lower abundance of helium, and the ring current producing hydrogen and oxygen ENAs with energies exceeding 10 keV \cite{ENA_From_LEO_INAFRoma}.\\
For this purpose, the SWEATERS (Space WEATher Ena Radiation SensorS) project \cite{EGU}, funded by the Italian Space Agency (ASI), aims to develop a novel instrument based on a Micro Pattern Gaseous Detector (MPGD), designed to operate in space and to be capable of measuring the energy and simultaneously reconstructing the trajectory of ENAs in the range 1-100 keV. A key requirement of the MPGD is to operate with low-pressure gas ($\mathcal{O}$\,(100\,mbar)) so that the particle range extends over several millimetres in the gas volume, which is essential for accurate track reconstruction.
An ultra-thin carbon foil or a 2D nanomaterial will be used as the detector entrance window to convert the ENAs into ions, enabling their detection in the gas volume of the MPGD \cite{Allegrini}.\\
The detector has been demonstrated to work well at pressures as low as 75 mbar, where it achieves 40\% FWHM energy resolution with X-rays at 5.9 keV emitted from a $^{55}$Fe source \cite{SWEATERS_MM_XRayChar}. It is well known that low-energy ions produce less ionization in the active medium of a detector compared to electrons of the same kinetic energy.  
This effect can be quantified by the ionization quenching factor (IQF), which is defined as the ratio of the energy deposited by the ion through ionization  to its kinetic energy.
For ENA detection, an accurate measurement of the IQF is therefore essential to reconstruct the ion kinetic energy from the ionization energy measured in the detector.\\
In this paper, we present measurements of the IQF for H$_2^+$ and He$^+$ ions with energies between 2.5 and 5 keV, interacting with an Ar/CO$_2$ gas mixture (volume ratio 93:7) at pressures below 150 mbar. From these measurements we derive for the first time the $W$ values for H$_2^+$ and He$^+$ in Ar in this energy range, which has not been previously studied for these specific projectile-target combinations. Section 2 provides an overview of the theoretical framework for the quenching factor, its connection to the $W$ value, and the methodology employed to measure the IQF. The experimental setup and the SWEATERS detector are described in section 3. X-ray detector calibration and data analysis procedures are described in section 4, while the results are presented in section 5.

\section{Quenching factor and W-value}
\subsection{Theoretical background}
Low energy ions lose energy in a medium through inelastic interactions with atomic electrons  and elastic scattering in the screened electric field of the nuclei. 
Electronic losses dominate for fast ions, while nuclear energy losses become significant as the ion slows down and dominate  for ion velocities smaller than the electron orbital velocity.
In the Lindhard theory \cite{Lindhard}, the mean energy transferred to electrons in the form of ionization and excitation ($\eta$) and atomic motion ($\nu$) is calculated by integrating the electronic and nuclear energy losses until an ion of kinetic energy $E_i$ stops. In the case of projectile and target with the same atomic (mass) number $Z$ ($A$), a parametrization for the 
quenching factors $q$, defined as the ratio of the average $\eta$ to the kinetic energy, is derived in \cite{Lindhard}
\begin{equation}
q = \frac{\bar{\eta}}{E_i} = \frac{E_i - \bar{\nu}}{E_i} =\frac{k g (\epsilon)}{1 + k g (\epsilon)}
\label{q_Lindhard}
\end{equation}
where $k = 0.133\, Z^{2/3} A^{-1/2}$ 
is a scaling factor for the electronic stopping power
function,
$g(\epsilon) = 3 \epsilon^{0.15} + 0.7 \epsilon^{0.6} + \epsilon$\, 
is an empirical fitting function that approximates the energy dependence of the electron stopping power, and $\epsilon$  a dimensionless reduced energy  \cite{Sorensen}.\\
For projectiles and targets with different atomic numbers, 
empirical formulae have been developed to calculate $q$, but these are generally limited to a small number of species and provide only rough agreement with the sparse experimental results available \cite{Akira}.
Therefore, additional measurements are needed especially at lower energies and with a wider variety of projectile-target combinations. \\
A practical way to evaluate the IQF is to define it as  \cite{Katsioulas}
\begin{equation}
\text{IQF} = \frac{E_d}{E}   
\label{eq:IQFdef1}
\end{equation}
where $E_d$ is
the energy deposited by a particle through ionization and excitation as it passes through a detector medium, and $E$ is the particle initial kinetic energy.
However, many detectors (such as those using a gaseous medium) are sensitive only to ionization, meaning that they measure only the number of electron-ion pairs generated along the particle track in the medium.
The number of electron-ion pairs $N_e$ produced by the complete stopping ($E_d = E_e$) of the electron in a gas is  $N_e = E_e/W_e$,  
where the $W$-value is the average energy to produce an electron-ion pair \cite{Knoll}. $W_e$ is larger than the first ionisation potential of the medium since a significant part of the energy is lost in the production of atomic or molecular excited states. 
$W_e$ does not depend on $E_e$ for electrons with $E_e > 1000$ eV. This value is approximately 26.4\,eV in pure argon \cite{ICRU_31}.\\
Unlike electrons, incident ions transfer a non-negligible fraction $\bar{\nu}$ of their kinetic energy $E_i$ through collisions with the gas nuclei.
As a result, only the fraction $\left(1-\frac{\bar{\nu}}{E_i}\right)$ of $E_i$ contributes to the ionisation and excitation of the gas atoms. The number of electron-ion pairs produced by an ion is given by
\begin{equation}
N_i =  \frac{\left(1-\frac{\bar{\nu}}{E_i}\right)}{W_e}\, E_i = \frac{E_i}{W_i}
\label{eq:Ni}
\end{equation}
where $W_i = \frac{W_e}{\left(1-\frac{\bar{\nu}}{E_i}\right)}\,$ is defined as the average energy spent by the ion to produce an electron-ion pair in the gas. 
Since part of the ion's energy is lost to nuclear interactions, $W_i$ depends on the ion kinetic energy $E_i$ and is greater than $W_e$, the corresponding value for electrons.\\
Using Eq.~\eqref{eq:IQFdef1} the IQF for ions can be written as 
\begin{equation}
\text{IQF} = \frac{E_d}{E_i} = \frac{N_i\, W_e}{E_i} = \frac{W_e}{W_i (E_i)} 
\label{eq:IQFdef2}
\end{equation}
where dependence of $W_i$ on the ion kinetic energy  is explicitly emphasised. 
Here, $E_d = N_i\, W_e$ represents the electron-equivalent ionization energy, determined by calibrating the detector with electrons \cite{Katsioulas}.
$E_d$ corresponds to the kinetic energy 
$\tilde{E}_e$ of an electron that produces in the medium a number of electron-ion pairs 
$\tilde{N}_e$ equal to 
the number $N_i$ produced by an ion with kinetic energy $E_i$. 
Therefore, from Eq.~\eqref{eq:IQFdef2}, an equivalent definition of the IQF is
\begin{equation}
\text{IQF} = \frac{W_e}{W_i (E_i)} = \frac{\tilde{E}_e}{E_i} 
\label{eq:IQFdef3}
\end{equation}\\
\subsection{Experimental procedure to measure the IQF}
In our experimental setup (Section \ref{sec:IQF_Apparatus}) we use an accelerated ion beam with selectable energy up to 5\,keV and a  radioactive source emitting X-rays of energy $E_\gamma = 5.9$ keV  for the detector calibration. 
To calculate the IQF with Eq.\eqref{eq:IQFdef3}, 
it is essential to estimate, for each ion energy $E_i$, the corresponding energy $\tilde{E}_e$ of the electrons producing the same amount of ionization as the ions. The values of
$\tilde{E}_e$ can be obtained
from the measurements taken with X-rays of energy $E_\gamma$.
The photoelectrons produced by photoelectric effect of X-rays in the gas have an energy approximately equal to $E_\gamma$, as the binding energies of  electrons in the atom are much lower than $E_\gamma$.
Since $W_e$ is constant at this energy, we can determine $\tilde{E}_e$, the kinetic energy of the electron that produces $\tilde{N}_e$ electron-ion pairs, by measuring the average number $N_\gamma$ of electron-ion pairs generated by a photoelectron in the gas as
\begin{equation}
\tilde{E}_e = \frac{\tilde{N}_e}{N_\gamma}E_\gamma
\end{equation}
Then Eq.~\eqref{eq:IQFdef3}  can be written as
\begin{equation}
\text{IQF} = \frac{\tilde{E}_e}{E_i} = \frac{E_\gamma}{E_i} \frac{\tilde{N}_e}{N_\gamma} = 
\frac{E_\gamma}{E_i} \frac{N_i}{N_\gamma}
\label{eq:IQFb}
\end{equation}
In this way we can calculate the IQF by measuring the detector response to ions with energy $E_i$ and X-rays with energy $E_\gamma$, which release different levels of ionisation in the gas, $N_i$ (which equals $\tilde{N}_e$ by the definition) and $N_\gamma$, respectively.\\
Moreover, the detector has to be calibrated to relate the pulse height $H$ (in ADC units) of the measured signals to the 
number $n$ of primary ionization electrons in the gas
\begin{equation}
H = F(E) = a + b\, Q = a + b M n
\label{equ:ReadoutChainResponse}
\end{equation}
where $M$ is the detector gain, $a$ and $b$ are parameters of the linear response $F$ of the readout electronics, and $Q = M\,n$ is the amplified charge in the detector. 
We measure signals
\begin{eqnarray}
H_i = a + bM N_i    \\
H_\gamma = a + bM N_\gamma
\end{eqnarray}
for the ions and the X-rays respectively. 
Inverting these relations and substituting in Eq.\eqref{eq:IQFb}, we obtain
\begin{equation}
\text{IQF} =  \frac{\left(H_i-a\right) E_\gamma}{\left(H_\gamma -a\right) E_i}
\label{equ:IQFFormulaWithXrays}
\end{equation}
that can be used to calculate the IQF by measuring the mean value of the pulse height distributions for the ion beam and X-rays, knowing the ion and X-ray energies and the parameter $a$ from the calibration. 
We notice that in this procedure it is not necessary to know the parameter $b$ and the gain $M$ of the detector,  since they cancel out in the ratio. 
\section{Experimental setup}
\label{sec:IQF_Apparatus}
\subsection{The ion beam facility}
\begin{figure}[H]
\centering
\includegraphics[width = 0.8\textwidth]{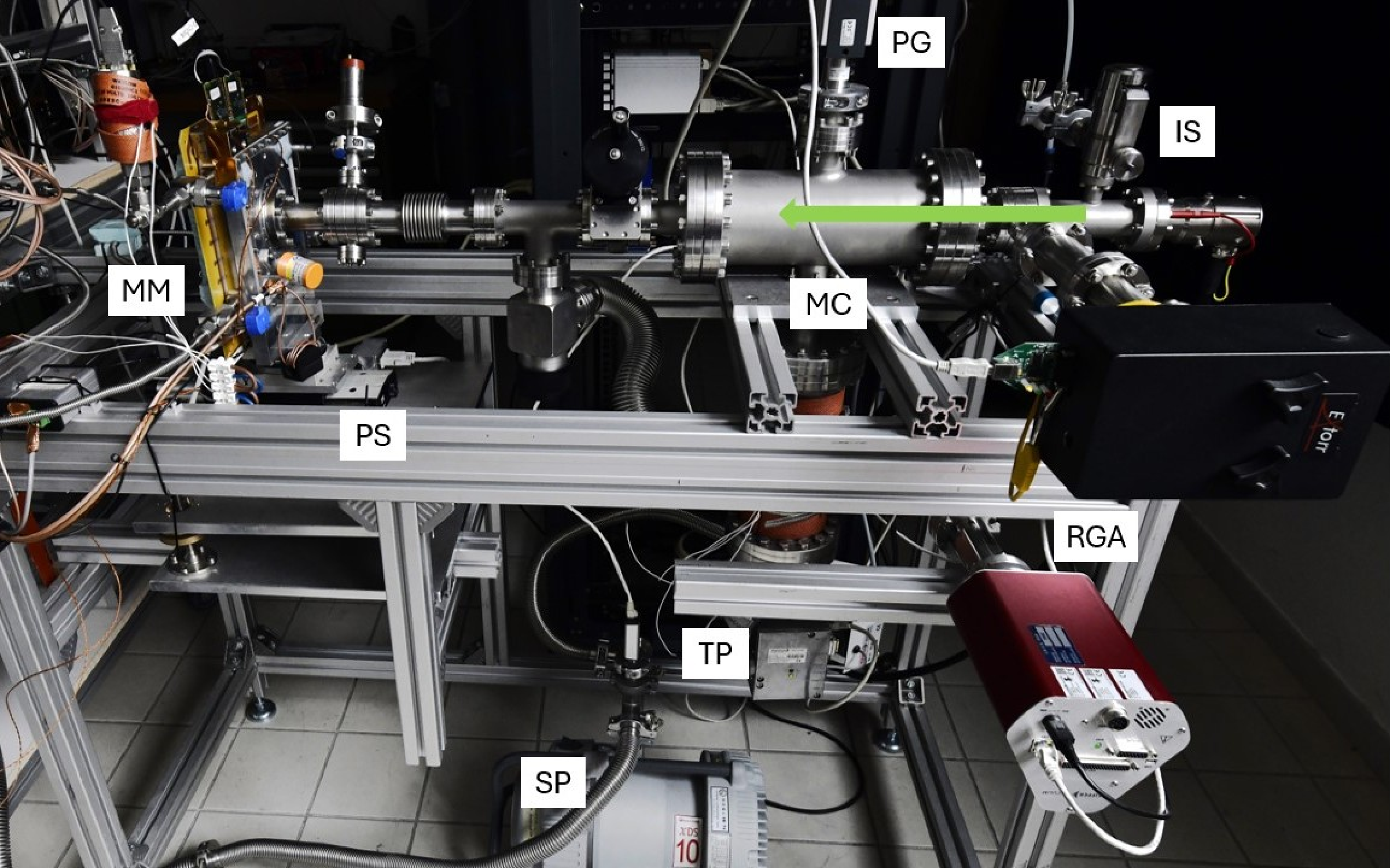}
\caption{
The Ion Beam Facility (IBF) directs an ion beam from the Ion Source (IS) toward the Main Chamber (MC), as indicated by a green arrow. The MC vacuum conditions are maintained using a Turbo Pump (TP) and a Scroll Pump (SP), and monitored with a Pressure Gauge (PG) and a Residual Gas Analyzer (RGA). 
The MM is mounted on a Positioning System (PS) that allows motion in all six degrees of freedom.}
\label{fig:IBF}
\end{figure}

The measurements in this paper were conducted at the Ion Beam Facility (IBF) at INFN Pisa (Fig.~\ref{fig:IBF}). The setup includes a sputter ion source (IS) that generates ion beams with energies between 0.2 and 5.0 keV and high energy stability (within 10 eV). 
High-purity helium and hydrogen are delivered to the IS via a contamination-controlled gas line that can be evacuated prior to use.
The IS connects to a main chamber maintained at ultra-high vacuum ($< 5\cdot10^{-8}$\,mbar) using turbo and scroll pumps. Vacuum conditions and gas purity are monitored using a pressure gauge and a residual gas analyzer.  
The detector is installed at the end of a  beam pipe connected to the main chamber and is mounted on an alignment system that combines manual and motorized stages, allowing for precise positioning \cite{SWEATERS2}. 
\subsection{The detector}
\label{subsec:detectorChar}
The detector is a resistive MICROMEGAS (MM) based on bulk technology \cite{Sauli, MM_Giomataris_FirstMM_1996}, fabricated by the CERN EP-DT-DD Micro Pattern Technology service \footnote{\url{https://ep-dep-dt.web.cern.ch/micro-pattern-technologies}}.
It features a 400-line-per-inch, 18-$\mu$m-diameter, nickel-woven wire mesh  suspended above the anode plane by insulating spacers (pillars) printed directly on the anode to ensure stable and uniform performance. 
The anode plane has a 2D readout structure consisting of two orthogonal layers of conductive strips embedded in an insulating substrate, enabling precise position reconstruction along both the x and y axes. There are 256 strips in each direction with a pitch of 400\,$\mu$m, covering an active area of 102.4$\times$102.4 mm$^2$. A thin layer of high-resistivity diamond-like carbon (DLC) coats the anode to provide effective spark protection while preserving spatial resolution through capacitive coupling with the underlying strips \cite{BulkMM}. The 192 $\mu$m amplification gap between the mesh and the anode is optimised to maximise charge amplification while preventing electrical breakdown at the operating pressures.
\begin{figure}[H]
\centering
\includegraphics[width=0.5\textwidth]{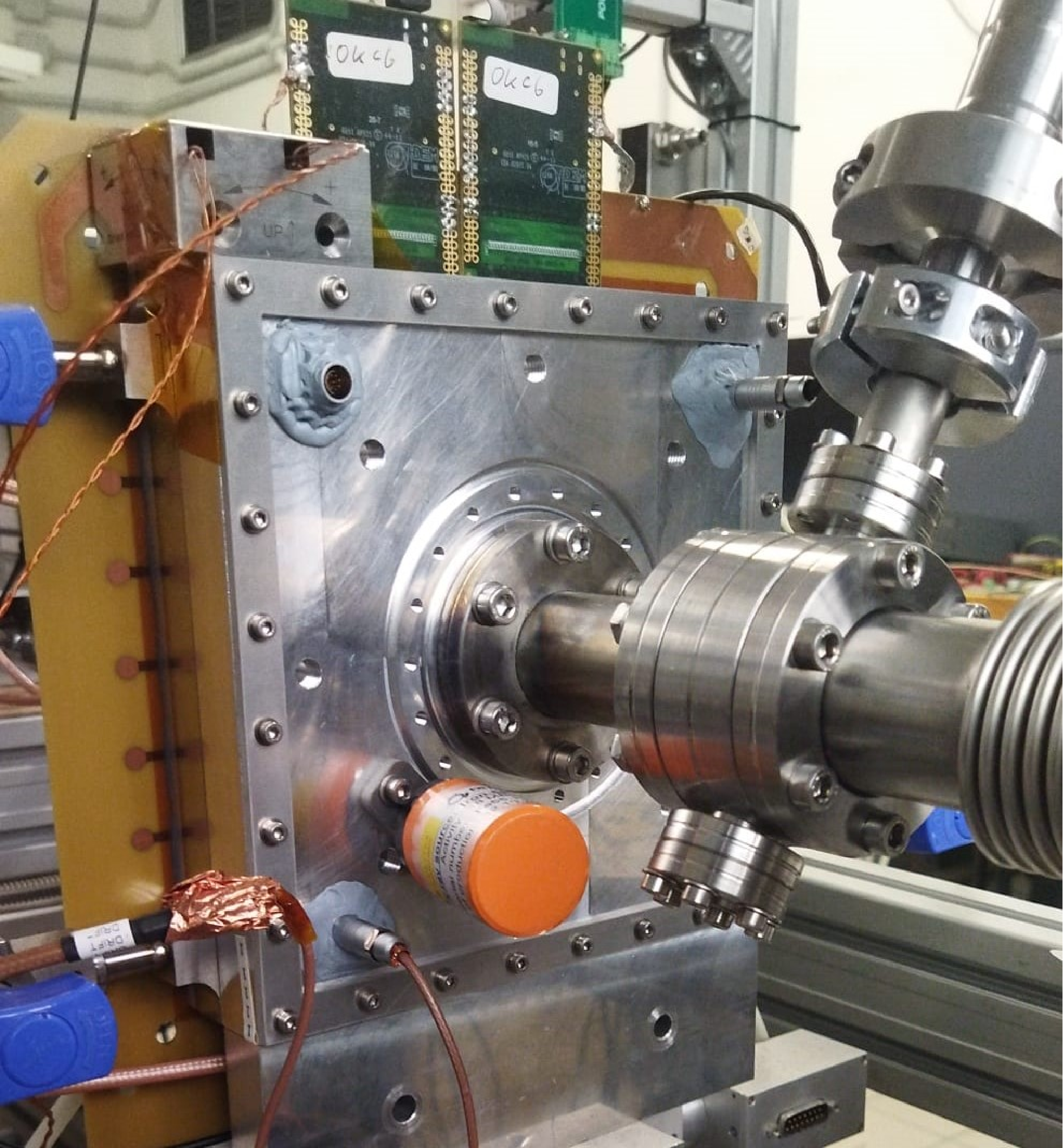}
\caption{Mechanical support frame of the MM detector 
connected to the beam pipe of the IBF. The $^{55}$Fe radioactive source (in its  orange protective container) is positioned in the lateral window. }
\label{fig:IQF_TopCover202406InstalledOnIBF}
\end{figure}
The MM is enclosed in a custom mechanical support structure consisting of a rear stiffener plate, a side frame and a gas-tight top cover (Fig.~\ref{fig:IQF_TopCover202406InstalledOnIBF}). The cathode, which is a square PCB with a side length of 100 mm, is mounted on the inner surface of the top cover, 20 mm from the anode to define the height of the drift volume. The top cover has two apertures: a central one containing a 5\,$\mu$m molybdenum pinhole through which the ion beam enters, and a nearby side window  with a diameter of 1 mm and sealed with a 100\,$\mu$m thick PEEK layer for positioning the X-ray calibration source.
\subsection{Electronic chain}
The voltages for the anode and cathode are delivered by a custom-designed, low-noise, high-voltage power supply.
Signals generated on the mesh electrode by electron avalanches are detected using a custom charge-sensitive preamplifier (CSP) with a dynamic range of 3-225 fC and an equivalent noise charge (ENC) of 0.650 fC \cite{MMCSA_SWEATERS}.
To suppress the low-frequency noise that arises when the MM operates under low-pressure conditions, a high-pass filter is connected in series with the CSP output. \\ 
The primary DAQ instrument is a computer-controlled DT5725SB CAEN digitizer\footnote{\url{https://www.caen.it/products/dt5725/}}, with 14-bit ADCs, which sample the CSP analogue waveforms at a rate of 250 MS/s for offline reconstruction and analysis. 
Furthermore, the CSP output signals are sent to low-threshold discriminator (CAEN N844\footnote{\url{https://www.caen.it/products/n844/}}) and  dual timer (CAEN N938\footnote{\url{https://www.caen.it/products/n93b/}}) NIM modules to  generate DAQ trigger signals with a fixed pulse width of 100\,ns and a 50\,$\mu$s hold-off period. This effectively prevents double counting of events that take longer to return to baseline.
For accurate trigger rate measurement, a Model TTi TF930 30 GHz counter\footnote{\url{https://www.aimtti.com/product-category/frequency-counters/aim-tf900series}} receives the NIM trigger signal.
\subsection{Gas distribution system}
The detector operates with a gas mixture of Ar/CO$_2$ in a volume ratio of 93\,\% and 7\,\%. 
A  gas distribution system has been designed to maintain an ultra-high purity gas flow with pressure stability within fractions of a mbar over long periods of time. Pressure\footnote{\url{https://www.nxp.com/part/MPX2200A}}, 
temperature and humidity\footnote{\url{https://sensirion.com/products/catalog/SHT35-DIS-B}} sensors
(with sensitivities of 0.5 mbar, 0.1\,$^{\circ}$C and 1\%, respectively)
are integrated both internally on the cathode PCB and externally on the MM, allowing real-time monitoring of the detector gas conditions and the laboratory environment.
The temperature of the detector is controlled to within 0.5\,$^\circ$C using a cooled liquid plate mounted on the bottom cover in conjunction with a dedicated cooling system.
\section{X-ray calibrations and data analysis}
\subsection{Pulse height spectra} 
\begin{figure}[h]
    \centering
    \includegraphics[width = 0.6\textwidth]{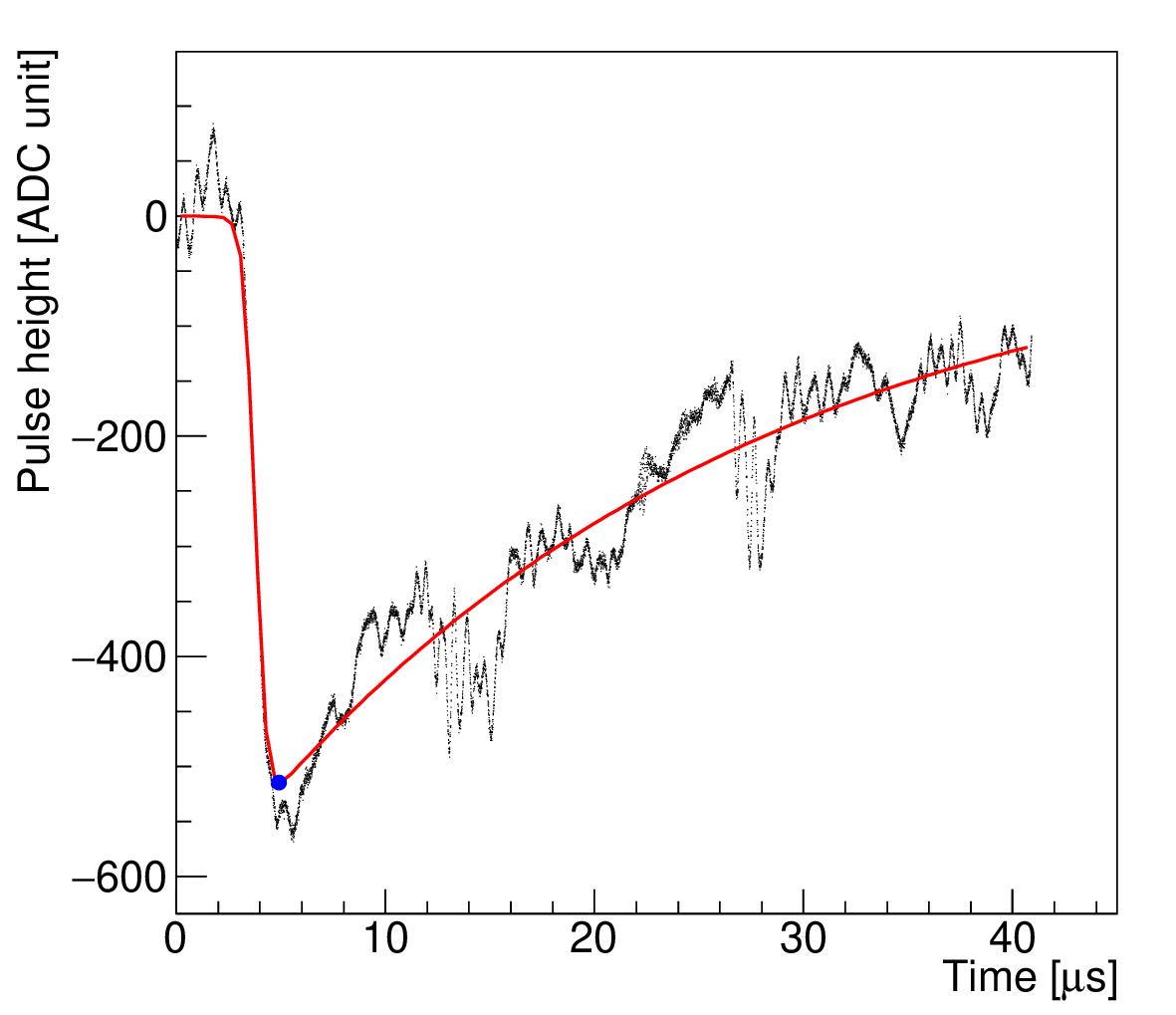} 
    \caption{Example of digitized waveform acquired when an  X-ray from a  $^{55}$Fe source is detected by the MM at a gas pressure of 75\,mbar. The red line represents the fitted function (Eq. \eqref{equ:FitFunction}). An estimate of the collected charge is given by the height of the fitted peak, indicated by the blue dot.}
    \label{fig:Fe55SignalExample_75mbar}
\end{figure}
\begin{figure}
    \begin{subfigure}{0.47\textwidth}
        \centering
        \includegraphics[width = \textwidth]{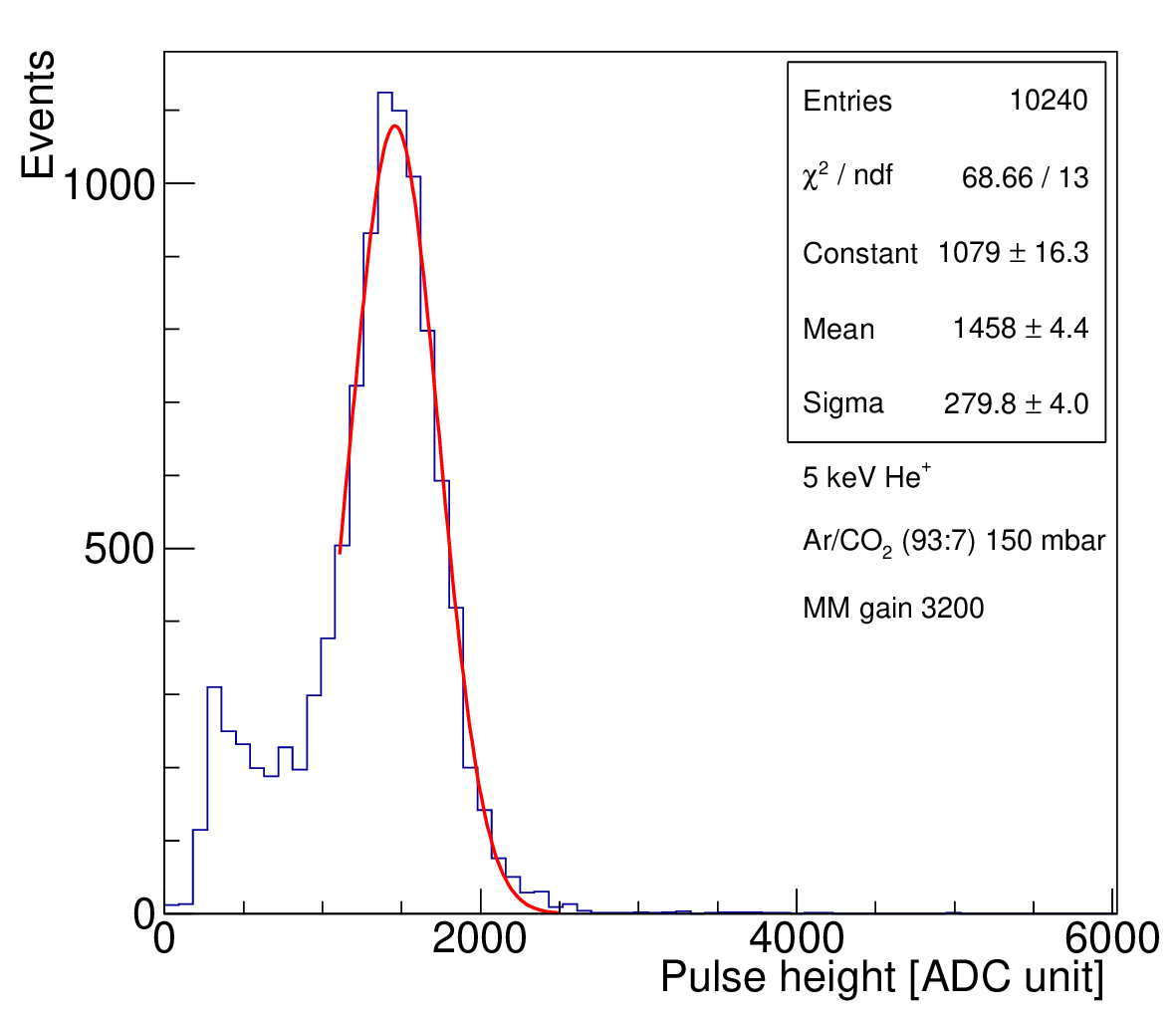} 
        \caption{}
        \label{fig:HeSpectrum150mbar}
    \end{subfigure}
    \hfill
    \begin{subfigure}{0.47\textwidth}
        \centering
        \includegraphics[width = \textwidth]{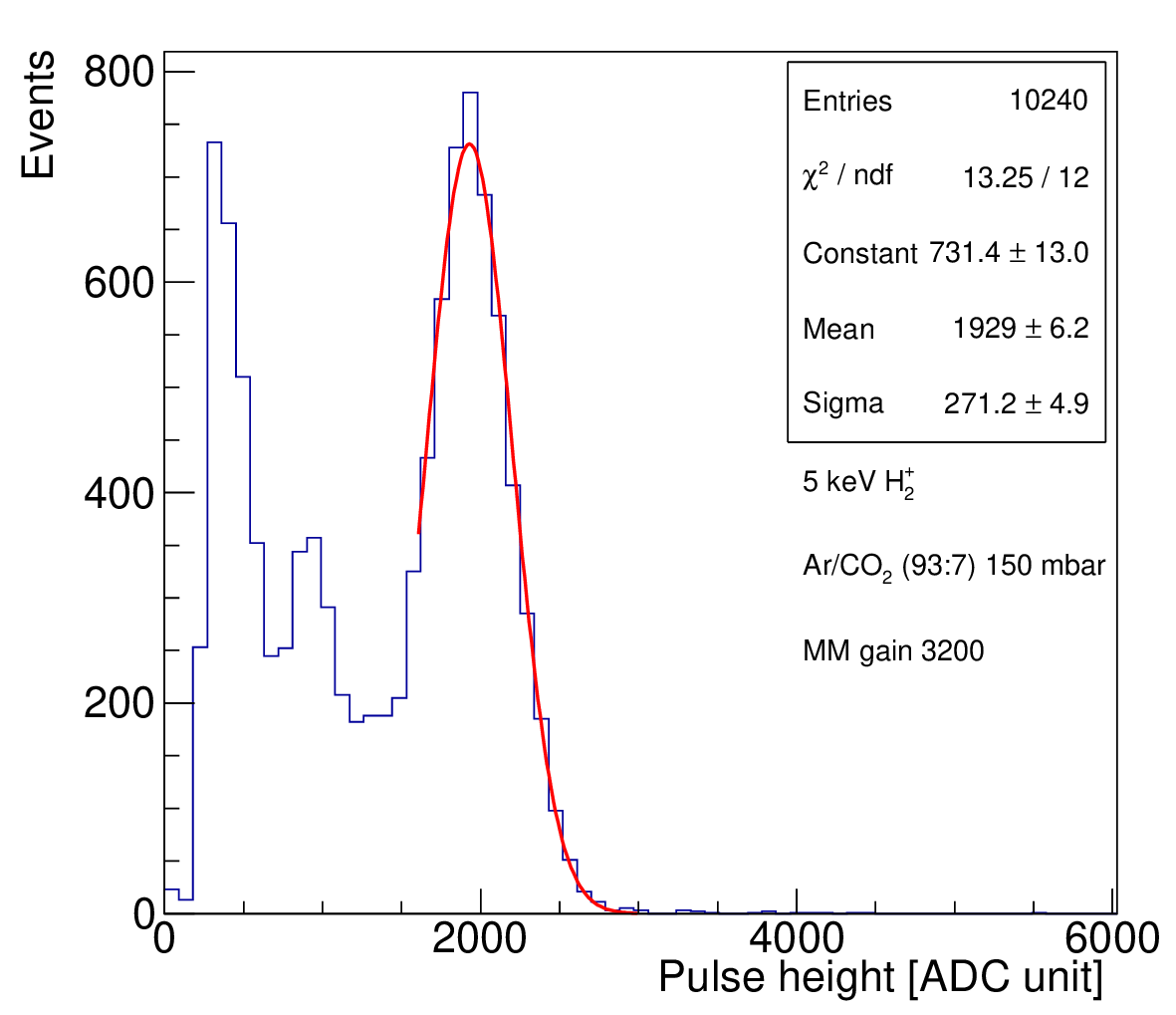} 
        \caption{}
        \label{fig:H2Spectrum150mbar}
    \end{subfigure}
    \caption{Pulse height distributions measured with the MM (gas pressure at 150 mbar) irradiated with 5\,keV He$^+$ (a) and H$_2^+$ (b) ions. The red lines represent Gaussian fits.}
\label{fig:PulseHightDistributions_Ions}
\end{figure}

The waveforms recorded by the digitizer, an examples of which is
shown in Fig.\,\ref{fig:Fe55SignalExample_75mbar} are converted 
into ROOT \cite{ROOT} files and analyzed offline with custom C++
codes. The pulse height is proportional to the total charge released in
the detector by the interacting particle. To extract the pulse height, each waveform is fitted to the function:
\begin{equation}
    f(t) = C\, \left[ \frac{1}{ \exp{\left(\frac{t-\mu}{\tau_f}\right)}+1}-1 \right]\, \exp{\left(\frac{
    2\mu - t}{\tau_r}\right)} 
    \label{equ:FitFunction}
\end{equation}
where the Fermi–Dirac term (enclosed in square brackets) models the fast charge collection on the MM mesh electrode, while the exponential tail accounts for the discharge of the amplifier. The function in Eq.\eqref{equ:FitFunction} is defined by four parameters: a normalization constant $C$; a shift parameter $\mu$, which corresponds to the trigger delay time; the time constant $\tau_f$ of the FD term; and the decay constant $\tau_r$ of the exponential component.  \\
Examples of pulse height distributions measured by the MM operating at a gas pressure of 150 mbar and irradiated with He$^+$ and H$_2^+$  ions at 5 keV are shown in Figs.~\ref{fig:HeSpectrum150mbar} and \ref{fig:H2Spectrum150mbar}. 
The main peaks (at higher ADC counts) are fitted to Gaussian functions to obtain the mean value $H_i$ of the MM response to each ion species $i$. 
It can be seen that the ion spectra show additional features beyond the main Gaussian peak.
Analysis using an RGA 
showed that heavier molecules (N$_2$, H$_2$O, CO$_2$) 
due to water contamination and surface outgassing accumulate in the ion source where they can be ionized and accelerated, contributing to the low-energy features in the distributions.
\subsection{Calibration of the electronic chain}
\begin{figure}
    \begin{subfigure}{0.47\textwidth}
        \centering
        \includegraphics[width = \textwidth]{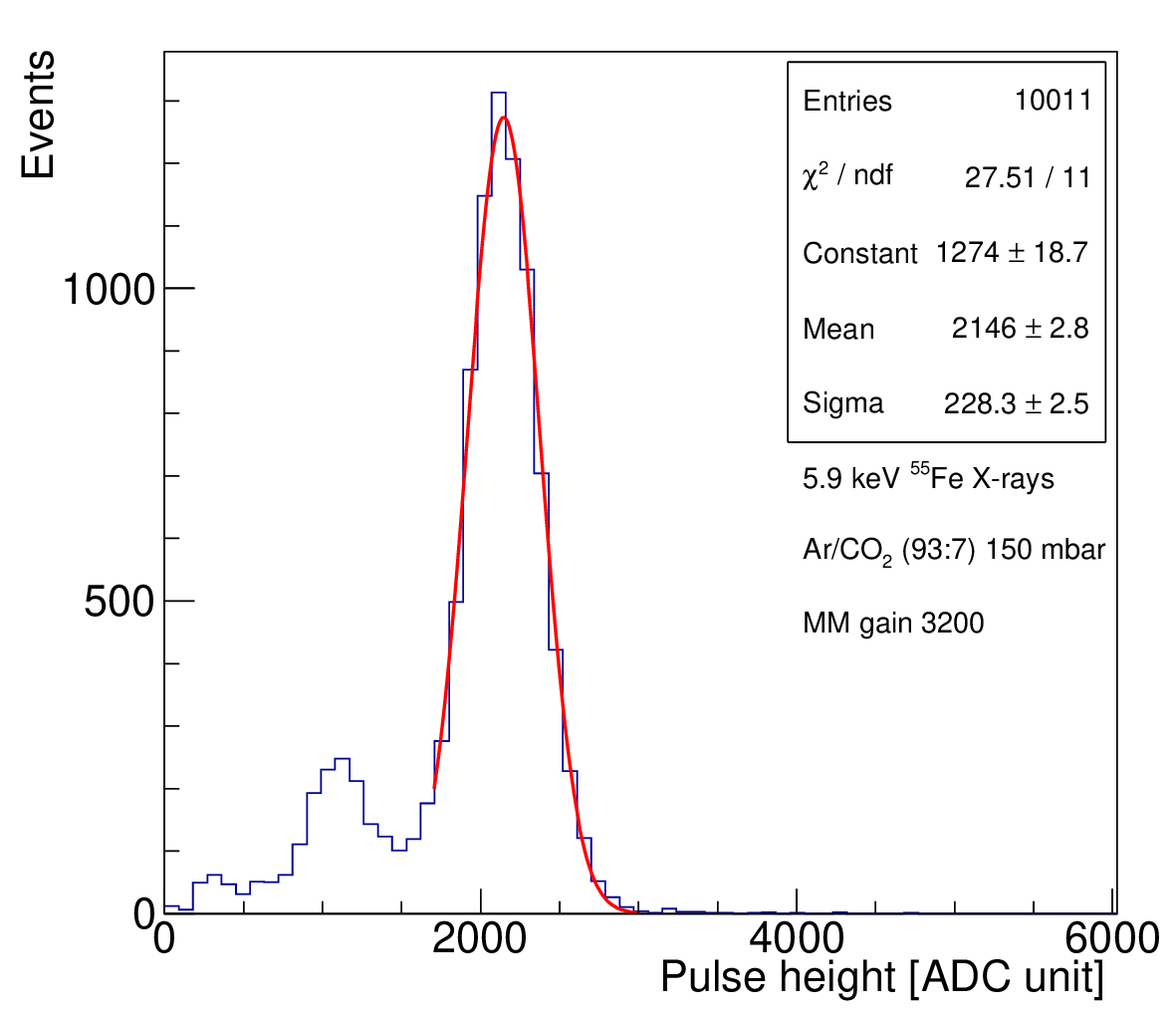} 
        \caption{}
        \label{fig:Fe55Spectrum150mbar}
    \end{subfigure}
    \begin{subfigure}{0.47\textwidth}
        \centering
        \includegraphics[width = \textwidth]{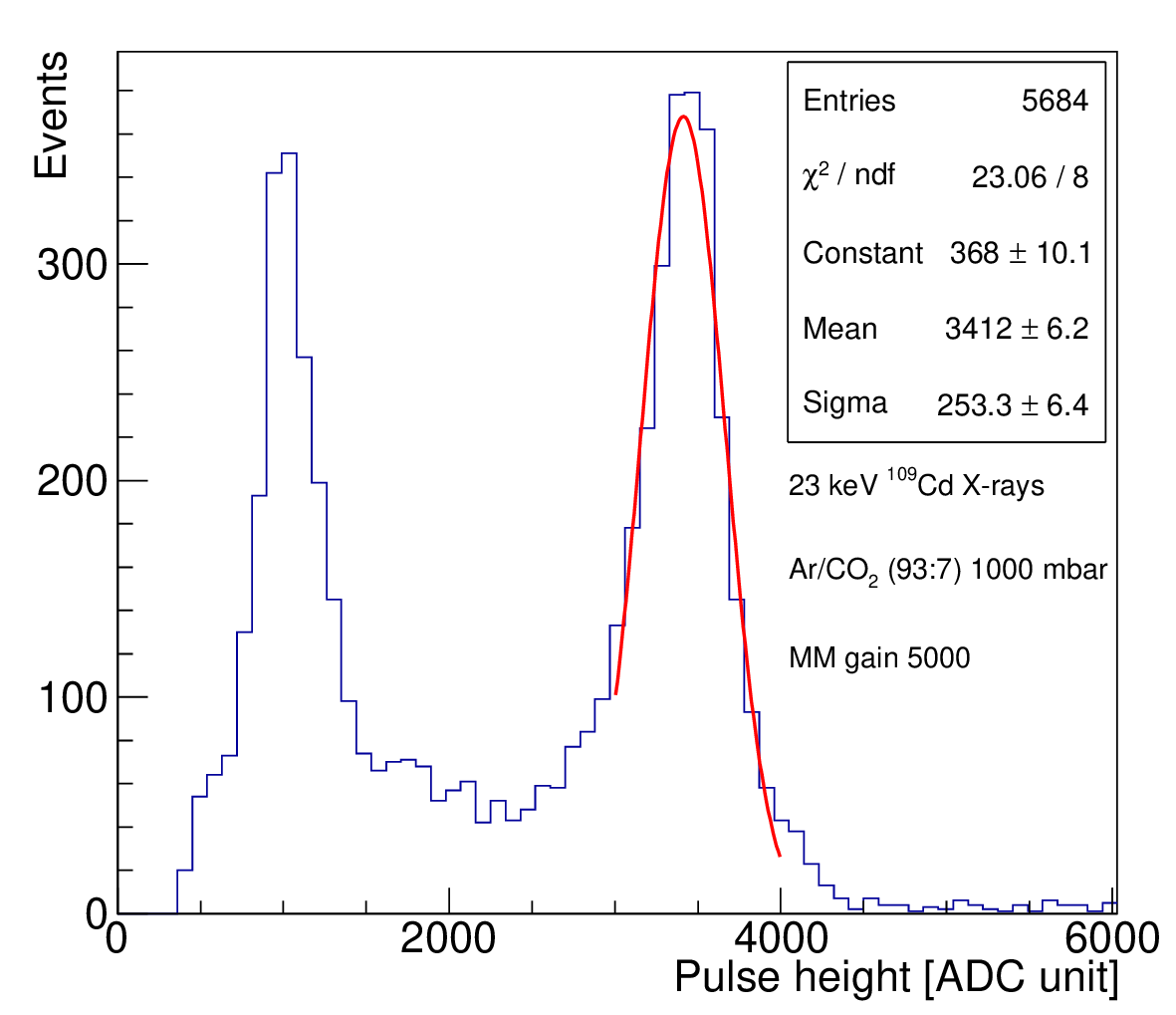} 
        \caption{}
        \label{fig:CdSpectrumNTP}
    \end{subfigure}
    \hfill
    \caption{Spectrum of $^{55}$Fe (a) and $^{109}$Cd (b)  measured with the MM held at 150 and 1000\, mbar respectively. 
    }
\label{fig:PulseHightDistributions_XRay}
\end{figure}

According to Eq.\,\eqref{equ:IQFFormulaWithXrays}, the value of the parameter $a$ is required for the IQF calculation. 
It can be determined by measuring the detector response to X-rays of two different energies and performing a linear interpolation of the data. For this purpose, $^{55}$Fe and $^{109}$Cd radioactive sources were used. These sources are
characterized by the emission of multiple K-shell X-rays by the daughter nuclei (Mn for Fe, Ag for Cd)
into which the parent nuclei decay by electron capture \cite{NIST}.
The corresponding spectra measured with the MM are shown in Figs.~\ref{fig:Fe55Spectrum150mbar} and \ref{fig:CdSpectrumNTP}
and exhibit several spectral features. 
The primary peak on the right is the MM response to the combined K-shell lines which cannot be resolved due to the 20\% FWHM energy resolution of the MM. The peak is centered on the average energy of the K-shell lines weighted by their respective emission probabilities (5.96\,keV  and 23.01\,keV  for $^{55}$Fe and $^{109}$Cd, respectively).
In Fig.~\ref{fig:Fe55Spectrum150mbar},  the Argon escape peak is visible at lower signal amplitudes,
resulting from the escape of fluorescence X-rays produced by Ar atoms. 
 In Fig.~\ref{fig:CdSpectrumNTP} the left peak is attributed to the K-shell emission of molybdenum, caused by 
 the photoelectric absorption of Cd X-rays by the Mo atoms of the pin-hole.  \\
The main peak in each spectrum is fitted to a Gaussian to determine the mean value in ADC of the MM response to X-rays of 5.96\,keV and 23.01\,keV, respectively.
The parameter $a$ is the intercept of the straight line passing through the two points of the graph  of the mean value in ADC counts as a function of the X-ray energy in keV. 
To verify the independence of $a$ from the detector operating point, data were obtained at two different pressures (500 and 1000 mbar), with the anode voltage adjusted at each pressure to obtain adequate gain. Measurements with the $^{109}$Cd source were not possible at lower pressures due to the very low interaction probability of X-rays in Ar gas. 
Combining all these measurements with the two X-ray sources gives a value of $a= (53.0\pm 5.5)$ ADC unit.\\
\subsection{Intercalibration of the response of X-ray irradiated MM in the central and lateral windows}
\begin{figure}[H]
\centering
\includegraphics[width = 0.7\textwidth]{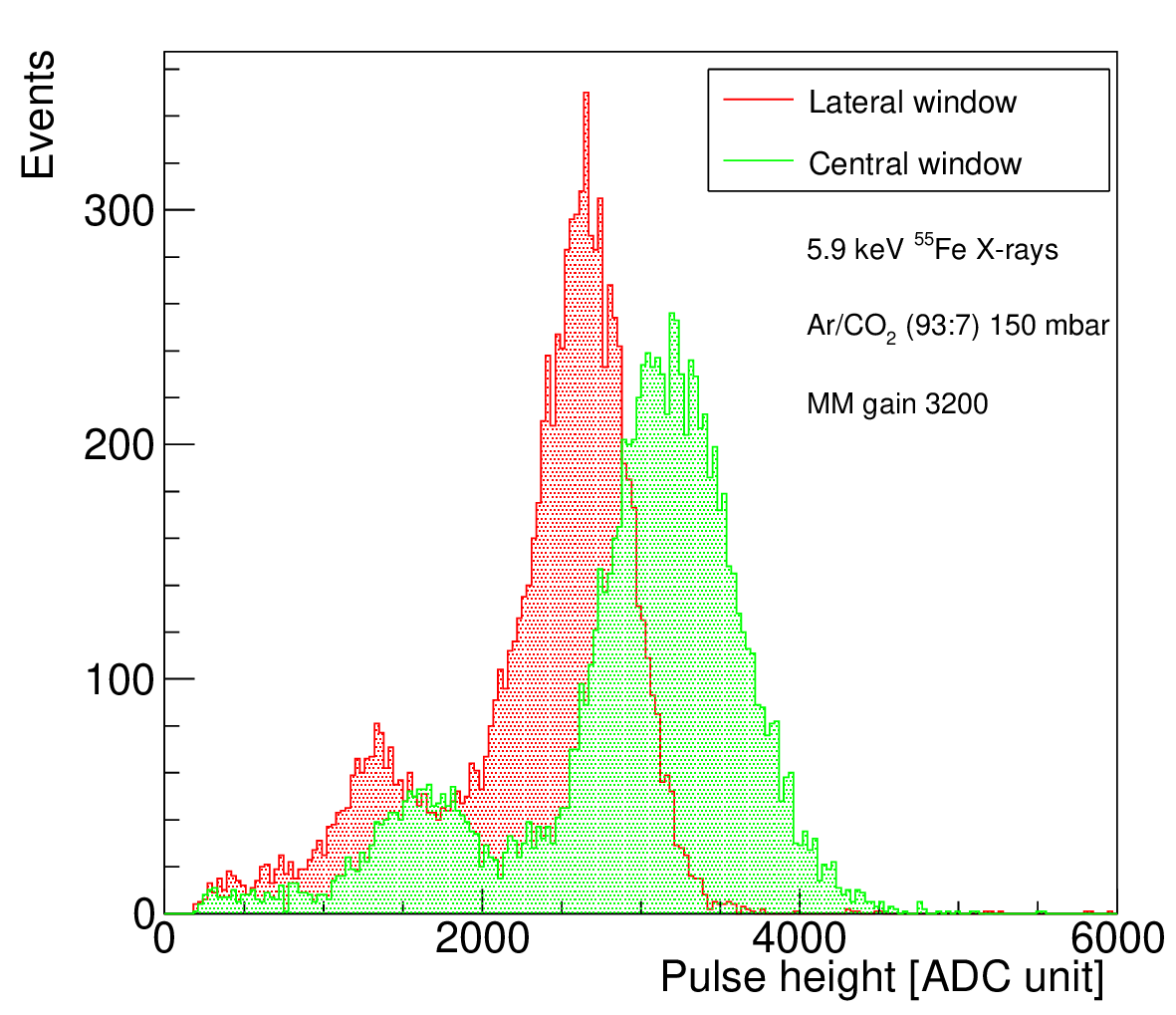} 
\caption{Pulse height distributions measured by the MM (gas pressure set to 150 mbar, anode voltage set to 310 V) with the $^{55}$Fe X-ray source placed in the central (green) and lateral window (red) of the detector cover. The value of $R$ for this working point is (0.163 $\pm$ 0.014) 
}
\label{fig:WinDiff150mbar_Spectra}
\end{figure}

In Eq.\,\eqref{equ:IQFFormulaWithXrays}, $H_\gamma$ is the mean signal of the pulse height distribution of the MM irradiated with X-rays emitted from an $^{55}$Fe source. 
As explained in Section~\ref{subsec:detectorChar}, when the beam is switched on, ions enter the MM through a pinhole in the center of the top cover, and 
the X-ray source is placed in a lateral aperture about 5\,cm away from the center. 
The response signal to X-rays is then formed in a region of the MM that is different from the region where the ions are measured.
The gain in the two regions could be different due to small differences in the avalanche electric field and gas pressure. 
In addition, a larger fraction of the X-ray ionisation could leak out of the detector if it is produced near the edge of the MM active region. 
The  result is a lower  $H_\gamma$ value which, if not corrected, 
could introduce systematic uncertainties in the measurement of the IQF. \\
To avoid that, we performed an intercalibration of the MM response to X-rays with the source placed in the central and lateral windows, when the beam pipe is not connected to the detector. 
A variation in the peak position was observed, as shown in Fig.~\ref{fig:WinDiff150mbar_Spectra}. 
This difference is quantified as the ratio   
\begin{equation}
R = \frac{H_\gamma^{\text{CW}}-H_\gamma^{\text{LW}}}{H_\gamma^{\text{CW}}}
\label{equ:RCorrFact}
\end{equation}
where $H_\gamma^{\text{CW}}$ and $H_\gamma^{\text{LW}}$ are the mean signals in  the central  and lateral windows, respectively. 
At a gas pressure of 150 mbar and an anode voltage of 310 V, the value of $R$ is (0.163 $\pm$ 0.014).
This correction factor $R$ is applied to the measurement of $H_\gamma^{LW}$  made when the beam pipe is connected to the central pinhole 
to estimate the X-ray signal $H_\gamma$ in Eq.\,\eqref{equ:IQFFormulaWithXrays}, which becomes
\begin{equation}
\text{IQF} =  \frac{\left(H_i - a\right)\, E_\gamma}{\left(H_\gamma - a\right)\, E_i}  = 
\frac{\left(H_i - a\right)\, E_\gamma}{\left[ \frac{H_\gamma^{\text{LW}}}{1-R}  - a\right]\, E_i} 
\label{equ:IQFFormula_Final2}
\end{equation}
\section{Experimental results}
\subsection{IQF measurements}
\begin{figure}[!h]
\centering
\includegraphics[width=0.7\linewidth]{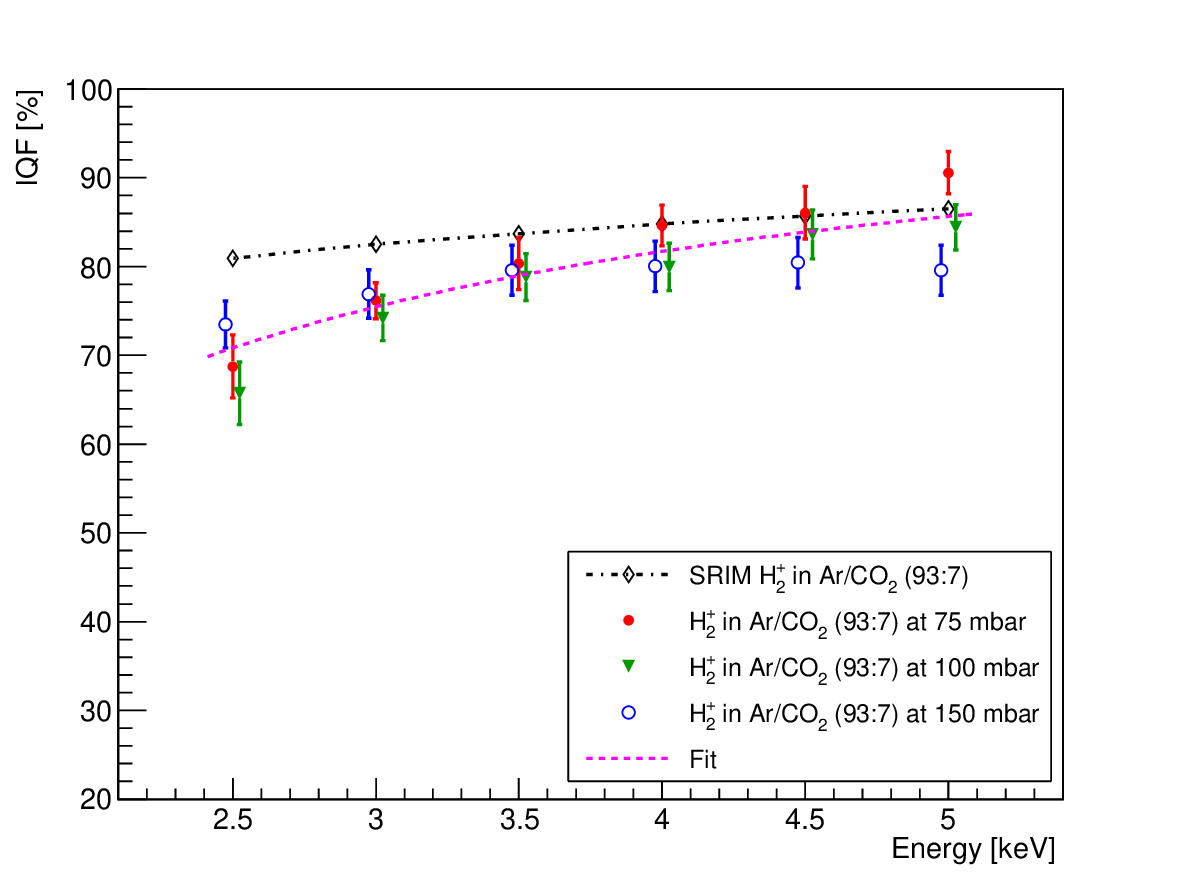}
\caption{IQF as a function of H$_2^+$ kinetic energy measured by the MM operating with Ar/CO$_2$ (93:7) at different pressures (75 mbar (filled circles), 100 mbar (triangles), 150 mbar (open circles)) 
and comparison with SRIM predictions (open diamonds). For clarity, the data points at 100 and 150 mbar are shifted horizontally. The magenta line is a fit with Eq.~\eqref{eq:IQF_vs_E} to the combined data sets at the three pressures.}
\label{fig:IQFH2_Press}
\end{figure}

\begin{figure}[!h]
\centering
\includegraphics[width=0.7\linewidth]{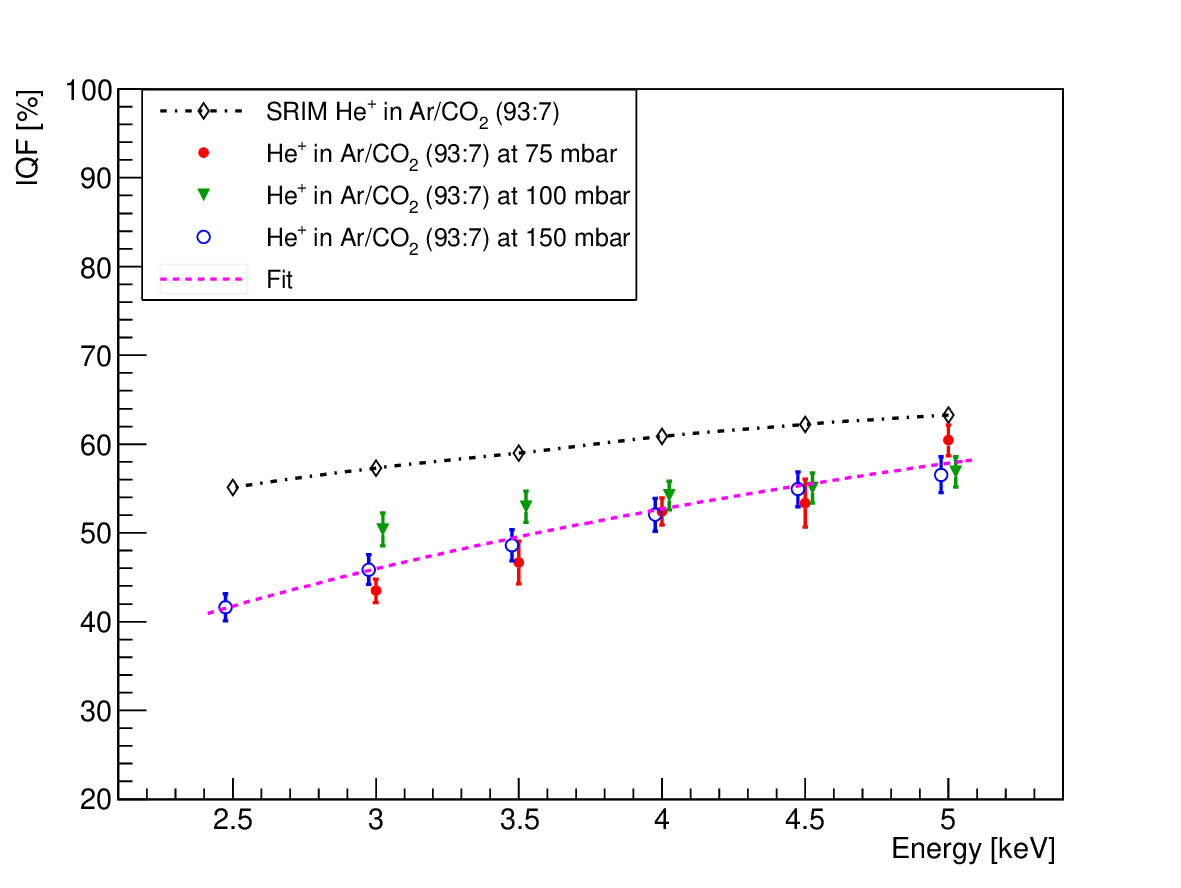}
\caption{IQF as a function of He$^+$ kinetic energy  measured by the MM operating with Ar/CO$_2$ (93:7) at different pressures (75 mbar (filled circles), 100 mbar (triangles), 150 mbar (open circles)) and comparison with SRIM predictions (open diamonds). For clarity, the data points at 100 and 150 mbar are shifted horizontally.
The magenta line is a fit with Eq.~\eqref{eq:IQF_vs_E} to the combined data sets at the three pressures. }
\label{fig:IQFHe_Press}
\end{figure}

The IQF was measured for He$^+$ and H$_2^+$ ions in the energy range of 2.5-5.0\,keV, with the MM operating at pressures of 75, 100 and 150\,mbar. The anode voltage was set to 254, 272 and 310 V respectively.  
The results in Figs.~\ref{fig:IQFH2_Press} and \ref{fig:IQFHe_Press} clearly show that there is a strong dependence 
of the IQF on the ion kinetic energy in the few keV region, and that the energy released in ionization  by He$^+$ is more quenched compared to that of H$_2^+$. \\
The errors of the data points are obtained by propagating in Eq.~\eqref{equ:IQFFormula_Final2} the errors of the parameters $a$ and $R$ and the error of the mean values $H_\gamma$ and $H_i$ from the Gaussian fits to the pulse height distributions. 
A systematic uncertainty of 2\% in the fitting procedure is estimated and added in quadrature to the error of the fitted mean  values. For each beam energy and gas pressure value, the measurements were repeated at least twice on different days and the results, found to be consistent within the errors, were averaged. \\
Notably, no clear trend of IQF with pressure was observed within the 75 mbar range examined. The data points at the three different pressures are consistent within the experimental uncertainties.\\
The combined data sets are fitted to a simplified function derived by Eq.~\eqref{q_Lindhard} \cite{Katsioulas}:
\begin{equation}
\text{IQF}(E) =\frac{E^{\alpha}}{\beta + E^{\alpha}}
\label{eq:IQF_vs_E}
\end{equation}
where $\alpha$ and $\beta$  are free parameters.
We obtain $\alpha_\text{H} = (1.30 \pm 0.17)$, $\beta_\text{H} = (1.35 \pm 0.29)$ 
and  $\alpha_\text{He} = (0.94 \pm 0.08)$, $\beta_\text{He} = (3.30 \pm 0.37)$,  
respectively. \\
The IQF of H$_2^+$ (He$^+$) increases rapidly from 70.8\% (41.7\%) at 2.5 keV to 85.6\% (57.9\%) at 5 keV.\\
We have compared the results with simulations based on SRIM (Stopping and Range of Ions in Matter) \cite{SRIM}, a software suite designed for calculating ion interactions and transport in matter.
The H$_2^+$ data agree with the SRIM data to within 3\% at $E\ge$4 keV, 
while at lower energies the relative difference increases up to 15\% at 2.5 keV.
For He$^+$ the relative difference between measurements and simulations is higher, increasing from 8\% at 5 keV to 24\% at 2.5 keV.
Similar discrepancies, with measured IQFs lower than those predicted by SRIM, have been reported for other gas mixtures \cite{Santos, Balogh}.
These differences may primarily be due to SRIM reliance on semi-empirical extrapolations of electronic and nuclear stopping powers at low energies, particularly below 10 keV, where experimental data are scarce \cite{SRIM}.  
In this regime, the electronic stopping power is estimated by subtracting a modeled nuclear stopping component from the measured total energy loss. Consequently, any inaccuracies in the nuclear stopping model directly propagate into the electronic stopping estimate, introducing systematic uncertainties into SRIM’s predictions \cite{Paul}. Further uncertainties arise from the SRIM simplified treatment of complex gas-phase ionization dynamics at low energies, including charge exchange, molecular excitation and dissociation \cite{Lan}. Nevertheless, SRIM simulations agree with our results in showing no pressure dependence of the IQF.
\begin{figure}[!h]
\centering
\includegraphics[width=0.7\linewidth]{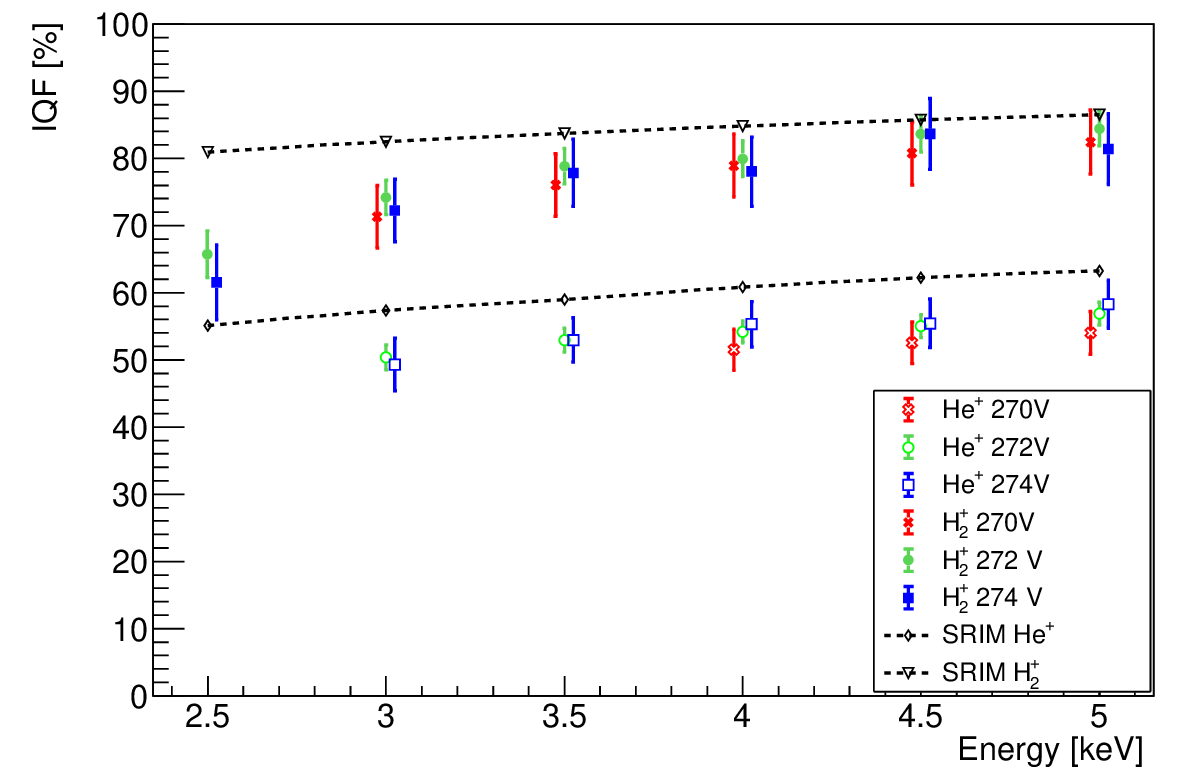}
\caption{IQF for H$_2^+$ and He$^+$ as a function of the ion kinetic energy  measured by the MM operating at 100 mbar and with different anode voltages. For clarity, the data points at 270 and 274 V anode voltage are shifted horizontally.}
\label{fig:IQFVoltage}
\end{figure}\\
The dependence of the IQF on the detector gain was also investigated at the fixed pressure of 100\,mbar by varying the anode voltage
around the reference value of 272 V, at which the gain is $\sim$1500.  At this very low pressure, the MM is very sensitive to small changes in the anode voltage \cite{SWEATERS_MM_XRayChar}: a change of $\pm$2\,V produces a gain variation of $\pm$50\%.
The measurements are consistent within the errors as shown in Fig.~\ref{fig:IQFVoltage}, confirming that the IQF does not depend 
on the gain of the MM, as expected. 
\subsection{$W$ values for H$_2^+$ and He$^+$ ions in Ar}
\begin{figure}[!h]
\centering
\includegraphics[width=0.8\linewidth]{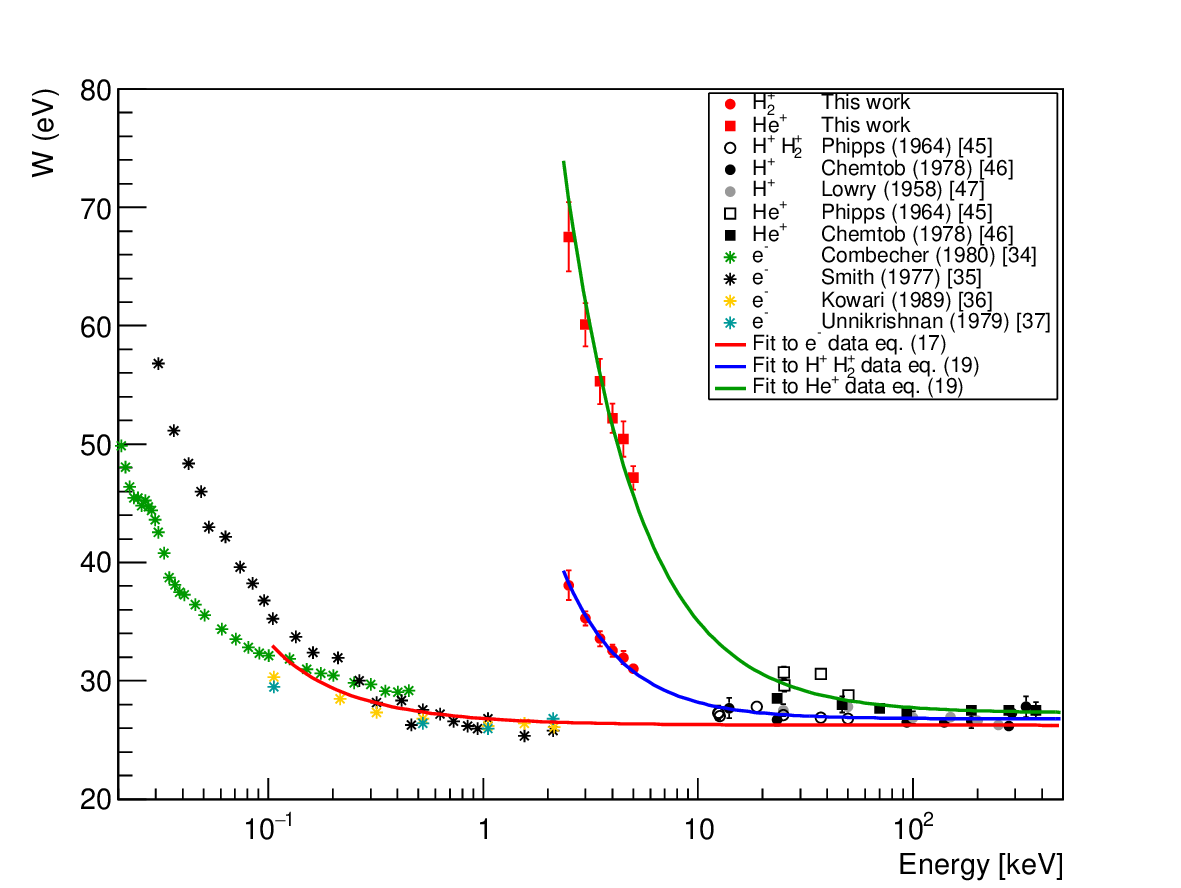}
\caption{W values of electrons (asterisks: green \cite{Wele1}, black \cite{Wele2}, yellow \cite{Wele3}, cyan \cite{Wele4}), H ions (red dots [this work], open circle \cite{W_Phipps}, black circles \cite{W_Chemtob}, gray circles \cite{W_Lowry}) and He$^+$ (red squares [this work], open squares \cite{W_Phipps}, black squares \cite{W_Chemtob}) in argon gas. 
The red line is a fit with Eq.~\eqref{eq:Wefit} to the electron data at energies $>$0.1 keV. 
The blue (green) line represents the fit with Eq.~\eqref{eq:Wifit} to the H (He) ion data.}
\label{fig:W}
\end{figure}

From Eq.~\eqref{eq:IQFdef2} the $W$ value at an ion energy $E_i$  can be calculated as 
\begin{equation}
W_i (E_i) = \frac{W_e}{\text{IQF}(E_i)}
\label{eq:IQF_W2}
\end{equation}
using the measured IQF and knowing the $W_e$ for electrons.
As shown in Fig.~\ref{fig:W}, $W_e$ increases with decreasing electron energy, 
due to the higher cross-section for excitation compared to ionisation at energies below 
a few hundred eV \cite{Katsioulas}.
This behaviour is usually described by the function \cite{W_IAEA}
\begin{equation}
W_{e}(E) = \frac{W_{e,a}}{1 - \frac{U}{E}}
\label{eq:Wefit}
\end{equation}
where $W_{ea}$ is the asymptotic value at high energies (much higher than the first ionization potential of the atom) and $U$ is a constant related to the mean energy of  sub-excitation electrons.
\begin{table}
\begin{center}
\begin{tabular}{|c|c|c|c|c|}
\hline
E [keV] & $W_{\text{H$_2^+$}}^{\text{mix}}$ [eV] & $W_{\text{He$^+$}}^{\text{mix}}$ [eV] & $W_{\text{H$_2^+$}}^{\text{Ar}}$ [eV] & $W_{\text{He$^+$}}^{\text{Ar}}$ [eV]     \\
\hline
2.50 & 38.3\,$\pm$\,1.1 & 64.8\,$\pm$\,2.7  & 38.1\,$\pm$\,1.3 & 67.5\,$\pm$\,2.9 \\ 
3.00 & 35.6\,$\pm$\,0.6 & 58.1\,$\pm$\,1.7  & 35.3\,$\pm$\,0.6 & 60.1\,$\pm$\,1.8\\   
3.50 & 33.9\,$\pm$\,0.6 & 53.8\,$\pm$\,1.7  & 33.6\,$\pm$\,0.6 & 55.3\,$\pm$\,1.9\\   
4.00 & 32.9\,$\pm$\,0.5 & 51.0\,$\pm$\,1.1  & 32.6\,$\pm$\,0.5 & 52.2\,$\pm$\,1.2\\  
4.50 & 32.3\,$\pm$\,0.5 & 49.4\,$\pm$\,1.3  & 32.0\,$\pm$\,0.6 & 50.4\,$\pm$\,1.5\\   
5.00 & 31.4\,$\pm$\,0.4 & 46.4\,$\pm$\,0.9  & 31.0\,$\pm$\,0.4 & 47.1\,$\pm$\,1.0\\         
\hline
\end{tabular}
\caption{$W$ values for H$_2^+$ and He$^+$ in Ar/CO$_2$ (93\%:7\%) mixture (second and third columns) and in pure Ar (fourth and fifth columns) as a function of the ion kinetic energy, calculated from the measured IQF. 
\label{tab:W}}
\end{center}
\end{table}
Fitting Eq.~\eqref{eq:Wefit} to a compilation of electron data in pure Ar gas \cite{Wele1, Wele2, Wele3, Wele4} 
yields $U= (17.7 \pm 0.9)$ eV and $W_{e,a} = (26.2 \pm 0.3)$ eV, which is in agreement with the value $(26.4 \pm 0.5)$ eV recommended by ICRU \cite{ICRU_31}.
It is clear from from Fig.~\ref{fig:W} that the asymptotic value $W_{ea}$ can be already used 
in the energy range of our measurements at a few keV. 
However, it has to be corrected for the fact that the MM does not work with pure Ar, 
but with an Ar mixture with 7\% CO$_2$.
In a regular gas mixture, $W_e^{\text{mix}}$ can be calculated as the weighted average of the $W_e$ values for the pure components \cite{W_IAEA}
\begin{equation}
W_e^{\text{mix}} = \frac{ W_e^{\text{Ar}}\,C_{\text{Ar}}\,\sigma_{\text{Ar}} + W_e^{\text{CO$_{2}$}}\,C_{\text{CO$_{2}$}}\,\sigma_{\text{CO$_{2}$}}}{C_{\text{Ar}}\,\sigma_{\text{Ar}} + C_{\text{CO$_{2}$}}\,\sigma_{\text{CO$_{2}$}}}
\label{eq_Wmix}
\end{equation}
where the weights are the product of the concentration fractions $C$ of each component by the its total ionisation cross section by electrons $\sigma$ \cite{XS, XS2}, evaluated at sufficiently high electron energy. In our case we calculate
$W_{e.a}^{\text{mix}} = (27.0 \pm 0.5)$ eV.\\
Using this latter value and the weighted average of the IQF values measured at the 
three different pressures for each ion energy, 
we have calculated the $W$ values of H$_2^+$ and He$^+$ in Ar/CO$_2$ (93\%:7\%), 
which are given in Table~\ref{tab:W} together with the derived $W$ values in pure Ar. 
These were calculated from $W$ in the gas mixture by inverting Eq.~\eqref{eq_Wmix}
and using the $W$ data for ions in CO$_2$ \cite{W_CO2, W_He_CO2}
and the total ionization cross sections of Ar and CO$_2$ by H$^+$ and He$^+$ ions reported in \cite{XS_p, XS_He}. \\
The $W$ values in pure Ar obtained in this work are shown in Fig.~\ref{fig:W}, where
we have also included all the available data on $W_{\text{H}}$ and $W_{\text{He}}$ in Ar at energies below 1 MeV. 
These are from measurements with accelerated H$^+$ and H$_2^+$ ions at energies between 12.5 and 374 keV 
and He$^+$ between 23.4 and 374 keV \cite{W_Phipps, W_Chemtob, W_Lowry}. 
It has been explained that there is no difference between the
$W$ values for H$^+$ and H$_2^+$ because the H$_2$ molecule dissociates in its initial collisions with the gas molecules \cite{W_Ruber}.
All these data cover an energy range where the $W$ values for H$_2^+$ and He$^+$ ions are already quite close to their asymptotic values. 
To the best of our knowledge, the new data presented in this paper represent the first measurements of $W_{\text{H}}$ and $W_{\text{He}}$ 
in Ar in the sub-10 keV energy region, where the strong energy dependence of the $W$ value  is instead evident.\\
The entire data collection can be fitted to a function which is the inverse of Eq.~\eqref{eq:IQF_vs_E} 
multiplied by the asymptotic value $W_{i,a}$
\begin{equation}
W_i(E) = \frac{\beta + E^{\alpha}}{E^{\alpha}}\, W_{i,a}
\label{eq:Wifit}
\end{equation}
yielding $W_{\text{H},a} = (26.8 \pm 0.1)$ eV and $W_{\text{He},a} = (27.3 \pm 0.2)$ eV.
\section{Conclusion}
We have measured the IQF of $H_2^+$ and He$^+$ ions in an Ar/CO$_2$ gas mixture at low pressure using a MM detector and a 
laboratory sputter ion source that accelerate ions between 2.5 and 5 keV. 
The results showed a strong dependence of the IQF on the ion kinetic energy in the few keV range, 
with no significant variation of the IQF observed with changes in gas pressure over the range studied.
Our measurements yielded lower IQF values than those predicted by SRIM simulations, with discrepancies increasing at lower ion energies, likely due to an overestimation of electronic stopping powers in SRIM, as reported in several studies.\\
Given the limitations of simulations and theoretical models at low energies, accurately calibrating the SWEATERS MM ionization response is crucial for its effective use as a spaceborne detector of ENAs in Earth’s magnetosphere.\\
Additionally, the measurements allowed us to derive the $W$ values for H$_2^+$ and He$^+$ in Ar 
in the previously unexplored sub-10 keV energy region. 
We observed a significant energy dependence, with $W$ values rapidly decreasing before 
reaching an asymptotic value at energies above 10 keV. 
These findings offer valuable insights into low-energy ion interactions in rarefied gases and 
can help improve the modeling of such interactions in experiments 
involving gaseous detectors, such as those for microdosimetry, 
direct dark matter searches, plasma monitoring, and space weather studies.
\section*{Acknowledgments}
This work has been supported by the ASI-INAF agreement “SWEATERS (Space WEATher Ena Radiation Sensors) – Phase B” n. 2024-7-HH.0 

We extend our gratitude to S. Bianucci and A. Soldani for their helpful contributions to the mechanical design, and to M. Ceccanti and A. Sardelli from INFN Sezione di Pisa for their exemplary technical expertise. Furthermore, we acknowledge our colleagues from the CERN RD51 collaboration for their significant contributions and constructive exchanges throughout the entire process of fine-tuning of our detectors.


\end{document}